# Secured Cryptographic Key Generation From Multimodal Biometrics: Feature Level Fusion of Fingerprint and Iris

A.Jagadeesan
Research scholar/Senior Lecturer/EIE
Bannari Amman Institute of Technology
Sathyamangalam-638 401, Tamil Nadu, India
.

Dr. K.Duraiswamy
Dean/Academic
K.S.Rangasamy College of Technology,
Tiruchengode – 637 209, Tamil Nadu, India
.

*Abstract*— **Human users have a tough time remembering long cryptographic keys. Hence, researchers, for so long, have been examining ways to utilize biometric features of the user instead of a memorable password or passphrase, in an effort to generate strong and repeatable cryptographic keys. Our objective is to incorporate the volatility of the user's biometric features into the generated key, so as to make the key unguessable to an attacker lacking significant knowledge of the user's biometrics. We go one step further trying to incorporate multiple biometric modalities into cryptographic key generation so as to provide better security. In this article, we propose an efficient approach based on multimodal biometrics (Iris and fingerprint) for generation of secure cryptographic key. The proposed approach is composed of three modules namely, 1) Feature extraction, 2) Multimodal biometric template generation and 3) Cryptographic key generation. Initially, the features, minutiae points and texture properties are extracted from the fingerprint and iris images respectively. Subsequently, the extracted features are fused together at the feature level to construct the multi-biometric template. Finally, a 256-bit secure cryptographic key is generated from the multi-biometric template. For experimentation, we have employed the fingerprint images obtained from publicly available sources and the iris images from CASIA Iris Database. The experimental results demonstrate the effectiveness of the proposed approach.**

## I. INTRODUCTION

The need for reliable user authentication techniques has increased in the wake of heightened concerns about security and rapid advancements in networking, communication and mobility [1]. Most authentication systems of today control access to computer systems or secured locations using passwords, but it has not been very resilient to attacks (can be broken or stolen). Therefore, biometrics has now become a practicable alternative to traditional identification methods in many application areas [23]. Biometrics, described as the science of recognizing an individual based on her physiological or behavioral traits, is beginning to gain acceptance as a legitimate method for determining an individual's identity [1]. Biometric technologies have showed significance in a range of security, access control and monitoring applications. The technologies are still new and rapidly evolving [2]. Biometric systems offer several advantages over traditional authentication methods, namely, 1) biometric information cannot be acquired by direct covert observation, 2) It is impossible to share and difficult to reproduce, 3) It enhances user convenience by alleviating the need to memorize long and random passwords, 4) It protects against repudiation by the user. In addition, biometrics provides the same level of security to all users unlike passwords and is highly resistant to brute force attacks [3]. Many biometric characteristics are being used today, including fingerprint, DNA, iris pattern, retina, ear, face, thermogram, gait, hand geometry, palm-vein pattern, keystroke dynamics, smell, signature, and voice [16, 17].

Biometric systems that use a single trait for recognition (i.e., unimodal biometric systems) are often affected by several practical problems like noisy sensor data, non-universality and/or lack of distinctiveness of the biometric trait, unacceptable error rates, and spoof attacks [4]. A probable improvement, multimodal biometric systems overcome some of these problems by consolidating the evidence obtained from different sources [5] [6]. Multimodal biometric system utilizes two or more individual modalities, e.g., face, gait, Iris and fingerprint, to improve the recognition accuracy of conventional unimodal methods. Using multiple biometric modalities has been shown to decrease error rates, by providing additional useful information to the classifier. Different features can be used by a single system or separate systems that can operate independently and their decisions may be combined [7]. The multimodal-based authentication can help the system in increasing the security and efficiency in contrast to unimodal biometric authentication, and it would be very difficult for an adversary to spoof the system because of two distinct biometrics traits [15].

Recently, multimodal biometrics fusion techniques have attracted much attention as the supplementary information between different modalities could improve the recognition performance. Many works have concentrated on this area [8-10]. In general, they can be classified into three categories: fusion at the feature level, fusion at the match level and fusion at the decision level [6] [11]. Fusion at the feature level involves the integration of feature sets corresponding to multiple modalities. Since the feature set contains richer information about the raw biometric data than the match score or the final decision, integration at this level is expected to provide better recognition results. However, fusion at this level is difficult to achieve in practice because of the following reasons: (i) the feature sets of multiple modalities may be incompatible (e.g., minutiae set of fingerprints and eigen-coefficients of face); (ii) the relationship between the feature spaces of different biometric systems may not be known; and (iii) concatenating two feature vectors may result in a feature vector with very large dimensionality leading to the `curse of dimensionality' problem [12].

One recent development, biometric cryptosystems [13] combine cryptography and biometrics to benefit from the strengths of both fields. In such systems, while cryptography provides high and adjustable security levels, biometrics brings in non-repudiation and eliminates the need to remember passwords or to carry tokens etc [14]. Of late, the enhanced performance of cryptographic key generated from biometrics in terms of security has obtained massive reputation amongst the researchers and experimenters [18] and in the recent past, researchers have attempted towards merging biometrics with cryptography in order to enhance overall security, by eliminating the necessity for key storage using passwords [19-22]. Even though it is highly infeasible to break cryptographic keys generated from biometrics, the attackers are still in with a possibility of sneaking through cryptographic attacks. One effective solution with added security will be the incorporation of multimodal biometrics into cryptographic key generation; so as to achieve incredible security against cryptographic attacks.

Here, we present a proficient approach for the secure cryptographic key generation based on multiple modalities namely, Iris and fingerprint. Initially, the fingerprint features (minutiae points) are extracted from the fingerprint image by means of segmentation, Orientation field estimation and morphological operators. Similarly, the texture features are extracted from the iris image using segmentation, estimation of iris boundary and Normalization. The two extracted features, minutiae points and iris texture are then fused at feature level to construct the multimodal biometric template. Fusion at the feature level is accomplished using the processes namely, shuffling, concatenation and merging. Finally, multi-biometric template obtained is used to generate the secure 256-bit cryptographic key that is capable of providing better user authentication and security.

The rest of the paper is organized as follows. A concise review of the researches related to the proposed approach is given in Section II. The proposed approach for generation of multimodal-based cryptographic key is illustrated in Section III. The results obtained on experimentation of the proposed approach are provided in Section IV. Finally, the conclusions are summed up in Section V.

## II. REVIEW OF RELATED RESEARCHES

A handful of researches are available in the literature for generating cryptographic keys from biometric modalities and multimodal biometrics based user authentication. Recently, developing approaches for cryptographic key generation from biometric features and authenticating users by combining multiple biometric modalities have received a great deal of attention among researchers. A brief review of some recent researches is presented here.

A realistic and safe way to incorporate the iris biometric into cryptographic applications has been presented by Feng Hao et al. [31]. A recurring binary string, called as a biometric key, was created reliably from genuine iris codes. The key was created from a subject's iris image with the help of auxiliary error-correction data, which does not disclose the key and can be stored in a tamper-resistant token, like a smart card. The reproduction of the key revolves on two aspects: the iris biometric and the token. They assessed the method using iris samples from 70 different eyes, with 10 samples from each eye. They identified that an error-free key have been created reliably from genuine iris codes with a 99.5 percent achievement rate. They created up to 140 bits of biometric key which is sufficient for a 128-bit AES. A technique that generates deterministic bit-sequences from the output of an iterative one-way transform using entropy based feature extraction process coupled with Reed-Solomon error correcting codes has been presented by B. Chen and V. Chandran [21]. The technique was assessed using 3D face data and was proved to reliably generate keys of suitable length for 128-bit Advanced Encryption Standard (AES).

A biometric-key generation scheme based on a randomized biometric helper has been presented by Beng.A et al. [42]. The technique involves a randomized feature discretization process and a code redundancy construction. The former allowed one to control the intra-class variations of biometric data to the minimal level and the latter reduced the errors even more. The randomized biometric helper confirmed that a biometric-key was simple to be revoked when the key was compromised. The projected technique was assessed in the context of face data based on a subset of the Facial Recognition Technology (FERET) database. Sanaul Hoque et al. [43] have presented the direct generation of the biometric keys from live biometrics, under certain conditions, by partitioning feature space into subspaces and partitioning these into cells, where each cell subspace contributes to the overall key generated. They assessed the presented technique on real biometric data, instead of both genuine samples and attempted imitations. Experimental results have proved the reliability in possible practical scenarios for this technique.

A lattice mapping based fuzzy commitment method for cryptographic key generation from biometric data has been presented by Gang Zheng et al. [44]. This technique generated high entropy keys and also concealed the original biometric data. This makes it impossible to recover the biometric data even when the stored information in the system was open to an attacker. Simulated results have proved that its authentication accuracy was on par to the k-nearest neighbor classification. Tianhao Zhang et al. [45] have presented a Geometry Preserving Projections (GPP) method for subspace selection. It is capable of discriminating different classes and conserving the intra-modal geometry of samples within an identical class. With GPP, they projected all raw biometric data from different identities and modalities onto a unified subspace, on which classification can be executed. Also, the training stage was performed after having a unified transformation matrix to project different modalities. Experimental results have proved the effectiveness of the presented GPP for individual recognition tasks.

A fusion architecture based on Bayesian belief networks has been presented by Donald E. Maurer and John P. Baker *et al.* [46]. The technique fully exploited the graphical structure of Bayes nets to define and explicitly model statistical dependencies between relevant variables: per sample measurements like, match scores and corresponding quality estimates and global decision variables. These statistical dependencies are in the form of conditional distributions which are modeled as Gaussian, gamma, log-normal or beta. Each model is determined by its mean and variance, thus considerably minimizing training data needs. Furthermore, they retrieved the information from lower quality measurements by conditioning decision variables on quality as well as match score instead of rejecting them out of hand. Another characteristic of the technique was a global quality measure intended to be used as a confidence estimate supporting decision making. Introductory studies using the architecture to fuse fingerprints and voice were accounted.

An efficient multimodal face and fingerprint biometrics authentication system on space-limited tokens, e.g. smart cards, driver license, and RFID cards has been proposed by Muhammad Khurram Khana and Jiashu Zhanga [47]. Fingerprint templates were encrypted and encoded/embedded inside the face images in a way that the features does not get changed significantly during encoding and decoding. Experimental and simulation results proved that the presented technique was a proficient and a cheap alternative to the multimodal biometrics authentication on space-limited tokens without downgrading the overall decoding and matching performance of the biometrics system. A class-dependence feature analysis technique based on Correlation Filter Bank (CFB) technique for efficient multimodal biometrics fusion at the feature level is presented by Yan Yan and Yu-Jin Zhang [48]. In CFB, by optimizing the overall original correlation outputs the unconstrained correlation filter trained for a specific modality. So, the variation between modalities has been considered and the useful information in various modalities is totally utilized. Previous experimental outcome on the fusion of face and palmprint biometrics proved the advantage of the technique.

An authentication method for a multimodal biometric system identification using two features i.e. face and palmprint has been presented by M.Nageshkumar *et al.* [24]. The technique was created for application where the training data includes a face and palmprint. Mixing the palmprint and face features has improved the robustness of the person authentication. The final assessment was done by fusion at matching score level architecture where features vectors were formed independently for query measures and are then evaluated to the enrolment template, which were saved during database preparation. Multimodal biometric system was expanded through fusion of face and palmprint recognition.

## III. Proposed Approach For Cryptographic Key Generation From Multimodal Biometrics

Multimodal biometric authentication has been more reliable and capable than knowledge-based (e.g. Password) and token-based (e.g. Key) techniques and has recently emerged as an attractive research area [24]. Several researchers [45-48] have successfully made use of multiple biometric traits for achieving user authentication. Multimodal biometrics was aimed at meeting the stringent performance requirements set by security-conscious customers. Some good advantages of multimodal biometrics are 1) improved accuracy 2) secondary means of enrollment and verification or identification in case sufficient data is not extracted from a given biometric sample and 3) ability to detect attempts to spoof biometric systems through non-live data sources such as fake fingers. Two important parameters that determine the effectiveness of the multimodal biometrics are choice of the biometric traits to be combined and the application area. The different biometric traits include fingerprint, face, iris, voice, hand geometry, palmprint and more. In the proposed approach, we integrate fingerprint and iris features for cryptographic key generation. The use of multimodal biometrics for key generation provides better security, as it is made difficult for an intruder to spool multiple biometric traits simultaneously. Moreover, the incorporation of biometrics into cryptography shuns the need to remember or carry long passwords or keys. The steps involved in the proposed multimodal-based approach for cryptographic key generation are,

1) Feature extraction from fingerprint.

2) Feature extraction from iris.

3) Fusion of fingerprint and iris features.

4) Generation of cryptographic key from fused features.

### A. Minutiae Points Extraction from Fingerprints

This sub-section describes the process of extracting the minutiae points from the fingerprint image. We chose fingerprint biometrics chiefly because of its two significant characteristics: uniqueness and permanence (ability to remain unchanged over the lifetime). A fingerprint can be described as a pattern of ridges and valleys found on the surface of a fingertip. The ridges of the finger form the so-called minutiae points: ridge endings (terminals of ridge lines) and ridge bifurcations (fork-like structures) [26]. These minutiae points serve as an important means of fingerprint recognition. The steps involved in the proposed approach for minutiae extraction are as follows,

*1) Preprocessing:* The fingerprint image is first preprocessed by using the following methods,

- Histogram Equalization

- Wiener Filtering

***Histogram Equalization:*** Histogram equalization (HE) is a very common technique for enhancing the contrast of an image. Here, the basic idea is to map the gray levels based on the probability distribution of the input gray levels. HE flattens and stretches the dynamic range of the image's histogram resulting in overall contrast improvement of the image [32].

HE transforms the intensity values of the image as given by the equation,

$$s_k = T(r_k) = \sum_{j=1}^{k} P_r(r_j) = \sum_{j=1}^{k} \frac{n_j}{n}$$

Where $s_k$ is the intensity value in the processed image corresponding to intensity $r_k$ in the input image, and $p_r(r_j) = 1,2,3 \ldots L$ is the input fingerprint image intensity level [33].

*Wiener filtering:* Wiener filtering improves the legibility of the fingerprint without altering its ridge structures [34]. The filter is based on local statistics estimated from a local neighborhood $\eta$ of size $3 \times 3$ of each pixel, and is given by the following equation:

$$w(n_1, n_2) = \mu + \frac{\sigma^2 - v^2}{\sigma^2} \left( I(n_1, n_2) - \mu \right)$$

where $v^2$ is the noise variance, $\mu$ and $\sigma^2$ are local mean and variance and $I$ represents the gray level intensity in $n_1, n_2 \in \eta$ [35].

*2) Segmentation:* The fingerprint image obtained after preprocessing is of high contrast and enhanced visibility. The next step is to segment the preprocessed fingerprint image. First, the fingerprint image is divided into non-overlapping blocks of size 16x16. Subsequently, the gradient of each block is calculated. The standard deviation of gradients in X and Y direction are then computed and summed. If the resultant value is greater than the threshold value the block is filled with ones, else the block is filled with zeros.

*3) Orientation Field Estimation:* A fingerprint orientation field is defined as the local orientation of the ridge-valley structures [27]. To obtain reliable ridge orientations, the most common approach is to go through the gradients of gray intensity. In the gradient-based methods, gradient vectors $[g_x, g_y]^T$ are first calculated by taking the partial derivatives of each pixel intensity in Cartesian coordinates. Traditional gradient-based methods divide the input fingerprint image into equal-sized blocks of $N \times N$ pixels, and average over each block independently [25] [26]. The direction of orientation field in a block is given by,

$$\theta_B = \frac{1}{2} a \tan \left( \frac{\sum_{i=1}^{N} \sum_{j=1}^{N} 2 g_x(i,j) g_y(i,j)}{\sum_{i=1}^{N} \sum_{j=1}^{N} g_x^2(i,j) - g_y^2(i,j)} \right) + \frac{\pi}{2}$$

Note that function $a \tan(\cdot)$ gives an angle value ranges in $\left( -\pi, \pi \right)$ which corresponds to the squared gradients, while $\theta_B$ is the desired orientation angle within $[0, \pi]$.

*4) Image Enhancement:* It would be desirable to enhance the fingerprint image further prior to minutiae extraction. The fingerprint image enhancement is achieved by using,

- Gaussian Low-Pass Filter
- Gabor Filter

***Gaussian Low-Pass Filter:*** The Gaussian low-pass filter is used as to blur an image. The Gaussian filter generates a `weighted average' of each pixel's neighborhood, with, the average weighted more towards the value of the central pixels. Because of this, gentler smoothing and edge preserving can be achieved. The Gaussian filter uses the following 2-D distribution as a point-spread function, and is achieved by the convolution [28].

$$G(x,y) = \left( \frac{1}{2\pi\sigma} \right)^2 \exp\left\{ \frac{(x^2 + y^2)}{2\sigma^2} \right\}$$

Where, $\sigma$ is the standard deviation of the distribution.

***Gabor Filter:*** Mostly used contextual filter [29] for fingerprint image enhancement is Gabor filter proposed by Hong, Wan, and Jain [30]. Gabor filters have both frequency-selective and orientation-selective properties and they also have optimal joint resolution in both spatial and frequency domains. The following equation shows the 2-Dimensional (2-D) Gabor filter form [29],

$$G(x,y,\theta,f_0) = \exp\left\{ -\frac{1}{2} \left( \frac{x_\theta^2}{\sigma_x^2} + \frac{y_\theta^2}{\sigma_y^2} \right) \right\} \cos(2\pi f_0 x_\theta),$$

$$\begin{bmatrix} x_\theta \\ y_\theta \end{bmatrix} = \begin{bmatrix} \sin\theta & \cos\theta \\ -\cos\theta & \sin\theta \end{bmatrix} \begin{bmatrix} x \\ y \end{bmatrix}$$

where $\theta$ is the orientation of the filter, $f_0$ is the ridge frequency, $[x_\theta, y_\theta]$ are the coordinates of $[x, y]$ after a clockwise rotation of the Cartesian axes by an angle of $(90° - \theta)$, and $\sigma_x$ and $\sigma_y$ are the standard deviations of the Gaussian envelope along the $x$-and $y$-axes, respectively.

*5) Minutiae extraction:* The process of minutiae point extraction is carried out in the enhanced fingerprint image. The steps involved in the extraction process are,

- Binarization
- Morphological Operators

***Binarization:*** Binarization is the process of converting a grey level image into a binary image. It improves the contrast between the ridges and valleys in a fingerprint image, and thereby facilitates the extraction of minutiae. The grey level value of each pixel in the enhanced image is examined in the binarization process. If the grey value is greater than the global threshold, then the pixel value is set to a binary value one; or else, it is set to zero. The output of binarization process is a binary image containing two levels of information, the foreground ridges and the background valleys. The minutiae extraction algorithms are good operating on binary images where there are only two levels of interest: the black pixels that denote ridges, and the white pixels that denote valleys.

***Morphological Operations:*** Following the binarization process, morphological operators are applied to the binarized fingerprint image. The objective of the morphological operations is to eliminate obstacles and noise from the image. Furthermore, the unnecessary spurs, bridges and line breaks are removed by these operators. The process of removal of redundant pixels till the ridges become one pixel wide is facilitated by ridge thinning. The Ridge thinning algorithm utilized for Minutiae points' extraction in the proposed approach has been employed by the authors of [36]. The image is first divided into two dissimilar subfields that resemble a checkerboard pattern. In the first sub iteration, the pixel p from the initial subfield is erased only when all three conditions, G1, G2, and G3 are satisfied. While, in the second sub iteration, the pixel p from the foremost subfield is erased when all three conditions, G1, G2, and G3' are satisfied.

**Condition G1:**

$$X_H(P) = 1$$

Where

$$X_H(P) = \sum_{i=1}^{4} b_i$$

$$b_i = \begin{cases} 1 \text{ if } x_{2i-1} = 0 \text{ and } (x_{2i} = 1 \text{ or } x_{2i+1} = 1) \\ 0 \quad \text{otherwise} \end{cases}$$

$x_1, x_2, ..., x_8$ are the values of the eight neighbors of $p$, starting with the east neighbor and numbered in counter-clockwise order.

**Condition G2:**

$$2 \le \min\{n_1(p), n_2(p)\} \le 3$$

where

$$n_1(p) = \sum_{k=1}^{4} x_{2k-1} \vee x_{2k}$$

$$n_2(p) = \sum_{k=1}^{4} x_{2k} \vee x_{2k+1}$$

**Condition G3:**

$$(x_2 \vee x_3 \vee \overline{x}_8) \wedge x_1 = 0$$

**Condition G3':**

$$(x_6 \vee x_7 \vee \overline{x}) \wedge x_5 = 0$$

The resultant fingerprint image produced by the morphological thinning algorithm composes of ridges each one pixel wide. This improves the visibility of the ridges and enables effortless and effortless of minutiae points.

*B. Feature Extraction from Iris*

The process of extracting features from the iris image is discussed in this sub-section. Iris recognition has been recognized as an effective means for providing user authentication. One important characteristic of the iris is that, it is so unique that no two irises are alike, even among identical twins, in the entire human population [37]. The human iris, an annular part between the pupil (generally, appearing black in an image) and the white sclera has an extraordinary structure and offers a plenty of interlacing minute characteristics such as freckles, coronas, stripes and more. These visible characteristics, which are generally called the texture of the iris, are unique to each subject [38]. The steps involved in the feature extraction process of the iris image are given below.

*1) Segmentation:* Iris segmentation is an essential module in iris recognition because it defines the effective image region used for subsequent processing such as feature extraction. Generally, the process of iris segmentation is composed of two steps 1) Estimation of iris boundary and 2) Noise removal.

***Estimation of iris boundary:*** For boundary estimation, the iris image is first fed to the canny algorithm which generates the edge map of the iris image. The detected edge map is then used to locate the exact boundary of pupil and iris using Hough transform.

- ***Canny edge detection***

The Canny edge detection operator was developed by John F. Canny in 1986. It uses a multi-stage algorithm to detect a wide range of edges in images. Canny edge detection starts with linear filtering to compute the gradient of the image intensity distribution function and ends with thinning and thresholding to obtain a binary map of edges. One significant feature of the Canny operator is its optimality in handling noisy images as the method bridges the gap between strong and weak edges of the image by connecting the weak edges in the output only if they are connected to strong edges. Therefore, the edges will probably be the actual ones. Hence compared to other edge detection methods, the canny operator is less fooled by spurious noise [39].

- ***Hough Transform***

The classical Hough transform was concerned with the identification of lines in the image, but later, the Hough transform has been extended to identify positions of arbitrary shapes, most commonly circles or ellipses. From the edge map obtained, votes are cast in Hough space for the parameters of circles passing through each edge point. These parameters are the centre coordinates x and y, and the radius r, which are able to define any circle according to the equation,

$$x^2 + y^2 = r^2$$

A maximum point in the Hough space will correspond to the radius and centre coordinates of the circle best defined by the edge points.

**Isolation of Eyelids and Eyelashes:** In general, the eyelids and eyelashes occlude the upper and lower parts of the iris region. In addition, specular reflections can occur within the iris region corrupting the iris pattern. The removal of such noises is also essential for obtaining reliable iris information.

• Eyelids are isolated by fitting a line to the upper and lower eyelid using the linear Hough transform. A second horizontal line is then drawn, which intersects with the first line at the iris edge that is closest to the pupil; the second horizontal line allows maximum isolation of eyelid region.

• The eyelashes are quite dark compared with the surrounding eyelid region. Therefore, thresholding is used to isolate eyelashes.

*2) Iris Normalization:* Once the iris image is efficiently localized, then the next step is to transform it into the rectangular sized fixed image. The transformation process is carried out using the Daugman's Rubber Sheet Model.

**Daugman's Rubber Sheet Model:** Normalization process involves unwrapping the iris and converting it into its polar equivalent. It is done using Daugman's Rubber sheet model [40] and is shown in figure.

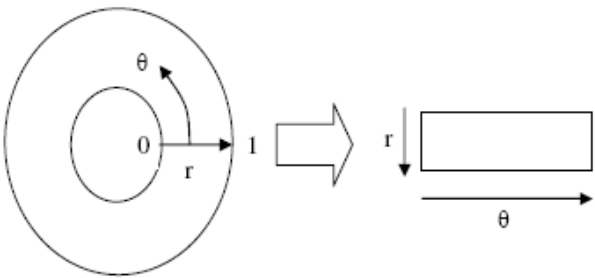

Figure 1. Daugman's Rubber Sheet Model

For every pixel in the iris, an equivalent position is found out on polar axes. The process comprises of two resolutions: Radial resolution, which is the number of data points in the radial direction and Angular resolution, which is the number of radial lines generated around iris region. Using the following equation, the iris region is transformed to a 2D array with horizontal dimensions of angular resolution and vertical dimension of radial resolution.

$$I[x(r,\theta), y(r,\theta)] \rightarrow I(r,\theta)$$

where, $I(x, y)$ is the iris region, $(x, y)$ and $(r, \theta)$ are the Cartesian and normalized polar coordinates respectively. The range of $\theta$ is $[0\ 2\pi]$ and $r$ is $[0\ 1]$. $x(r,\theta)$ and $y(r,\theta)$ are defined as linear combinations set of pupil boundary points. The formulas given in the following equations perform the transformation,

$$x(r,\theta) = (1-r)x_p(\theta) + x_i(\theta)$$

$$y(r,\theta) = (1-r)y_p(\theta) + y_i(\theta)$$

$$x_p(\theta) = x_{p0}(\theta) + r_p Cos(\theta)$$

$$y_p(\theta) = y_{p0}(\theta) + r_p Sin(\theta)$$

$$x_i(\theta) = x_{i0}(\theta) + r_i Cos(\theta)$$

$$y_i(\theta) = y_{i0}(\theta) + r_i Sin(\theta)$$

where $(x_p, y_p)$ and $(x_i, y_i)$ are the coordinates on the pupil and iris boundaries along the $\theta$ direction. $(x_{p0}, y_{p0})$, $(x_{i0}, y_{i0})$ are the coordinates of pupil and iris centers [39].

*3) Extraction of iris texture:* The normalized 2D form image is broken up into 1D signal, and these signals are used to convolve with 1D Gabor wavelets. The frequency response of a Log-Gabor filter is given as,

$$G(f) = \exp\left(\frac{-(\log(f/f_0))^2}{2(\log(\sigma/f_0))^2}\right)$$

Where $f_0$ represents the centre frequency, and $\sigma$ gives the bandwidth of the filter [41].

The Log-Gabor filter outputs the biometric feature (texture properties) of the iris.

*C. Fusion of Fingerprint and Iris Features*

We have at hand two sets of features namely, 1) Fingerprint features and 2) Iris features. The next step is to fuse the two sets of features at the feature level to obtain a multimodal biometric template that can perform biometric authentication.

*Feature Representation*: Fingerprint - Each minutiae point extracted from a fingerprint image is represented as $(x, y)$ coordinates. Here, we store those extracted minutiae points in two different vectors: Vector $F_1$ contains all the $x$ co-ordinate values and Vector $F_2$ contains all the $y$ co-ordinate values.

$$F_1 = [x_1\ x_2\ x_3 \ldots x_n]\ ; \ |F_1| = n$$

$$F_2 = [y_1\ y_2\ y_3 \ldots y_n]\ ; \ |F_2| = n$$

Iris - The texture properties obtained from the log-gabor filter are complex numbers $(a + ib)$. Similar to fingerprint representation, we also store the iris texture features in two different vectors: Vector $I_1$ contains the real part of the complex numbers and Vector $I_2$ contains the imaginary part of the complex numbers.

$$I_1 = [a_1 \ a_2 \ a_3 \dots a_m] \ ; \ |I_1| = m$$

$$I_2 = [b_1 \ b_2 \ b_3 \dots b_m] \ ; \ |I_2| = m$$

Thereby, the input to the fusion process (multimodal biometric generation) will be four vectors $F_1, F_2, I_1$ and $I_2$. The fusion process results with the multimodal biometric template. The steps involved in fusion of biometric feature vectors are as follows.

*1) Shuffling of individual feature vectors:* The first step in the fusion process is the shuffling of each of the individual feature vectors $F_1, F_2, I_1$ and $I_2$. The steps involved in the shuffling of vector $F_1$ are,

   i.   A random vector $R$ of size $F_1$ is generated. The random vector $R$ is controlled by the seed value.

   ii.   For shuffling the $i^{th}$ component of fingerprint feature vector $F_1$,

   a)   The $i^{th}$ component of the random vector $R$ is multiplied with a large integer value.

   b)   The product value obtained is modulo operated with the size of the fingerprint feature vector $F_1$.

   c)   The resultant value is the index say '$j$' to be interchanged with. The components in the $i^{th}$ and $j^{th}$ indexes are interchanged.

   iii.   Step (ii) is repeated for every component of $F_1$. The shuffled vector $F_1$ is represented as $S_1$.

The above process is repeated for every other vectors $F_2, I_1$ and $I_2$ with $S_1 \ S_2$ and $S_3$ as random vectors respectively, where $S_2$ is shuffled $F_2$ and $S_3$ is shuffled $I_1$. The shuffling process results with four vectors $S_1, S_2, S_3$ and $S_4$.

*2) Concatenation of shuffled feature vectors:* The next step is to concatenate the shuffled vectors process $S_1, S_2, S_3$ and $S_4$. Here, we concatenate the shuffled fingerprints $S_1$ and $S_2$ with the shuffled iris features $S_3$ and $S_4$ respectively. The concatenation of the vectors $S_1$ and $S_3$ is carried out as follows:

   i.   A vector $M_1$ of size $|S_1| + |S_3|$ is created and its first $|S_3|$ values are filled with $S_3$.

   ii.   For every component $S_1$,

   a) The corresponding indexed component of $M_1$ say '$t$' is chosen.

   b) Logical right shift operation is carried in $M_1$ from index '$t$'.

   c) The component of $S_1$ is inserted into the emptied $t^{th}$ index of $M_1$.

The aforesaid process is carried out between shuffled vectors $S_2$ and $S_4$ to form vector $M_2$. Thereby, the concatenation process results with two vectors $M_1$ and $M_2$.

*3) Merging of the concatenated feature vectors:* The last step in generating the multimodal biometric template $B_T$ is the merging of two vectors $M_1$ and $M_2$. The steps involved in the merging process is as follows.

   i.   For every component of $M_1$ and $M_2$,

   a.   The components $M_{11}$ and $M_{21}$ are converted into their binary form.

   b.   Binary *NOR* operation is performed between the components $M_{11}$ and $M_{21}$.

   c.   The resultant binary value is then converted back into decimal form.

   ii.   These decimal values are stored in the vector $B_T$, which serves multimodal biometric template.

*D. Generation of Cryptographic Key from Fused Features*

The final step of the proposed approach is the generation of the k-bit cryptographic key from multimodal biometric template $B_T$. The template vector $B_T$ can be represented as,

$$B_T = [b_{T_1} \ b_{T_2} \ b_{T_3} \dots b_{T_h}]$$

The set of distinct components in the template vector $B_T$ are identified and are stored in another vector $U_{BT}$.

$$U_{BT} = [u_1 \ u_2 \ u_3 \cdots u_d] \ ; \ |U_{BT}| \le |B_T|$$

The vector $U_{BT}$ is then resized to $k$ components suitable for generating the *k*-bit key. The resize procedure employed in the proposed approach,

$$B = \begin{cases} [u_1 \ u_2 \ \cdots \ u_k] & ; if \ |U_{BT}| > k \\ [u_1 \ u_2 \ \cdots \ u_d] << u_i; \ d+1 \ge i \ge k & ; if \ |U_{BT}| < k \end{cases}$$

Where, $u_i = \dfrac{1}{d} \sum\limits_{j=1}^{d} u_j$

Finally, the key $K_B$ is generated from the vector B,

$$K_B << B_i \bmod 2, \ i = 1,2,3 \dots k$$

## IV. EXPERIMENTAL RESULTS

The experimental results of the proposed approach have been presented in this section. The proposed approach is implemented in Matlab (Matlab7.4). We have tested the proposed approach with different sets of fingerprint and iris images corresponding to individuals. The fingerprint images employed in the proposed approach have been collected from publicly available databases. The input fingerprint image, the extracted minutiae points and the intermediate results of the proposed approach are shown in figure 2. For iris feature extraction, we use iris images obtained from CASIA Iris Image Database collected by Institute of Automation, Chinese Academy of Science. The input iris image, the normalized iris image and the intermediate results of the proposed approach are portrayed in figure 3. Finally, the generated 256-bit cryptographic key obtained from the proposed approach is depicted in figure 4.

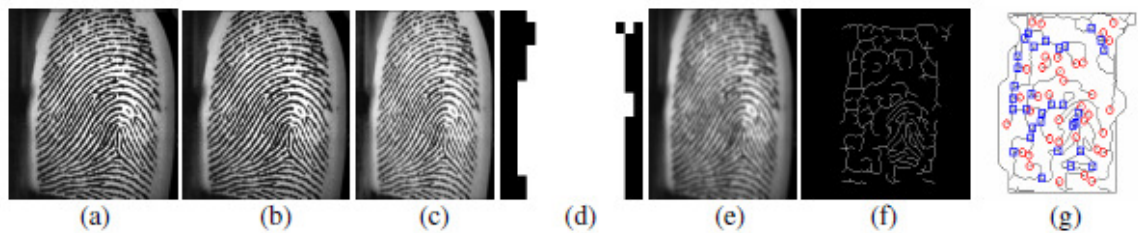

Figure 2. (a) Input fingerprint image (b) Histogram Equalized Image (c) Wiener Filtered Image (d) Segmented Image (e) Enhanced image (f) Morphological Processed Image (g) Fingerprint image with Minutiae points

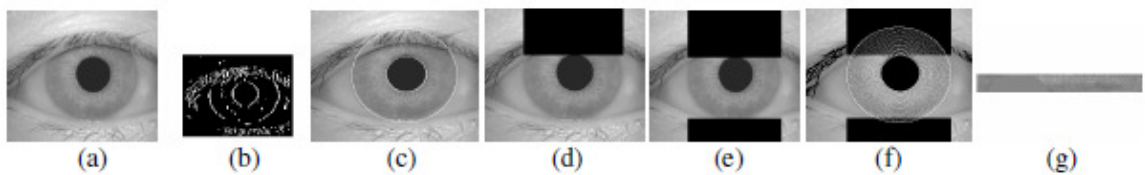

Figure 3. (a) Input Iris image (b) Edge detected image (c) Located pupil and iris boundary (d) Detected top eyelid region (e) Detected top and bottom eyelid region (f) Segmented Iris image (g) Normalized iris image

```
0101010101010001010101011010100101100010010001010101011001000100000100010100
1001110110101001000000010000010000010101011000000101000000100001000101010011001
00110100001001100000000001010010010000000100000000101000000010000100001000000010
```

Figure 4. Generated 256 bit key

## V. CONCLUSION

In this paper, we have attempted to generate a secure cryptographic key by incorporating multiple biometrics modalities of human being, so as to provide better security. An efficient approach for generation of secure cryptographic key based on multimodal biometrics (Iris and fingerprint) has been presented in this paper. The proposed approach has composed of three modules namely, 1) Feature extraction, 2) Multimodal biometric template generation and 3) Cryptographic key generation. Firstly, the features, minutiae points and texture properties have been extracted from the fingerprint and iris images respectively. Then, the extracted features have been combined together at the feature level to obtain the multi-biometric template. Lastly, a 256-bit secure cryptographic key has been generated from the multi-biometric template. For experimentation, we have employed the fingerprint images obtained from publicly available sources and the iris images from CASIA Iris Database. The experimental results have demonstrated the efficiency of the proposed approach to produce user-specific strong cryptographic keys.

## REFERENCES

[1] Arun Ross and Anil K. Jain, "Multimodal Biometrics: An Overview", in proceedings of the 12th European Signal Processing Conference, pp. 1221-1224, 2004.

[2] Richard A. Wasniowski, "Using Data Fusion for Biometric Verification", in Proceedings of World Academy of Science, Engineering and Technology, vol. 5, April 2005.

[3] Parvathi Ambalakat, "Security of Biometric Authentication Systems", in proceedings of 21st Computer Science Seminar, 2005.

[4] A.K. Jain and A. Ross, "Multi-biometric systems: special issue on multimodal interfaces that flex, adapt, and persist", Communications of the ACM, vol. 47, no. 1, pp. 34–40, 2004.

[5] L. Hong, A.K. Jain and S. Pankanti, "Can multibiometrics improve performance?", in Proceedings of IEEE Workshop on Automatic Identification Advanced Technologies, pp. 59–64, NJ, USA, 1999.

[6] Anil Jain, Karthik Nandakumar and Arun Ross, "Score normalization in multimodal biometric systems", Pattern Recognition, vol. 38, pp. 2270 – 2285, 2005.

[7] Eren Camlikaya, Alisher Kholmatov and Berrin Yanikoglu, "Multi-biometric Templates Using Fingerprint and Voice", Biometric technology for human identification, Vol. 6944, no5, pp: 1-9, Orlando FL, 2008.

[8] R. Wang and B. Bhanu, "Performance prediction for multimodal biometrics", In Proceedings of the IEEE International Conference on Pattern Recognition, pp. 586-589, 2006.

[9] X. Jing, Y. Yao, D. Zhang, J. Yang, and M. Li. "Face and palm print pixel level fusion and Kernel DCV-RBF classifier for small sample biometric recognition", Pattern Recognition, vol. 40, no.11, pp. 3209-3224, 2007.

[10] T. Zhang, X. Li, D. Tao, and J. Yang, "Multi-modal biometrics using geometry preserving projections", Pattern Recognition, vol. 41, no. 3, pp. 805-813, 2008.

[11] Yan Yan and Yu-Jin Zhang, "Multimodal Biometrics Fusion Using Correlation Filter Bank", in proceedings of 19th International Conference on Pattern Recognition, pp. 1-4, Tampa, FL, 2008.

[12] Arun Ross and Rohin Govindarajan, "Feature Level Fusion in Biometric Systems", in proceedings of Biometric Consortium Conference (BCC), September 2004.

[13] Umut Uludag, Sharath Pankanti, Salil Prabhakar, Anil K.Jain, "Biometric Cryptosystems Issues and Challenges", in Proceedings of the IEEE, vol. 92, pp. 948-960, 2004.

[14] P.Arul, Dr.A.Shanmugam, "Generate a Key for AES Using Biometric for VOIP Network Security", Journal of Theoretical and Applied Information Technology, vol. 5, no.2, 2009.

[15] Muhammad Khurram Khan and Jiashu Zhang, "Multimodal face and fingerprint biometrics authentication on space-limited tokens", Neurocomputing, vol. 71, pp. 3026-3031, August 2008.

[16] Kornelije Rabuzin and Miroslav Baca and Mirko Malekovic, "A Multimodal Biometric System Implemented within an Active Database Management System", Journal of software, vol. 2, no. 4, October 2007.

[17] M Baca and K. Rabuzin, "Biometrics in Network Security", in Proceedings of the XXVIII International Convention MIPRO 2005, pp. 205-210 , Rijeka,2005.

[18] N. Lalithamani and K.P. Soman, "Irrevocable Cryptographic Key Generation from Cancelable Fingerprint Templates: An Enhanced and Effective Scheme", European Journal of Scientific Research, vol.31, no.3, pp.372-387, 2009.

[19] A. Goh and D.C.L. Ngo, "Computation of cryptographic keys from face biometrics", International Federation for Information Processing 2003, Springer-Verlag, LNCS 2828, pp. 1–13, 2003.

[20] F. Hao, C.W. Chan, "Private Key generation from on-line handwritten signatures", Information Management & Computer Security, vol. 10, no. 2, pp. 159–164, 2002.

[21] Chen, B. and Chandran, V., "Biometric Based Cryptographic Key Generation from Faces", in proceedings of 9th Biennial Conference of the Australian Pattern Recognition Society on Digital Image Computing Techniques and Applications, pp. 394 - 401, December 2007.

[22] N. Lalithamani and Dr. K.P. Soman, "An Effective Scheme for Generating Irrevocable Cryptographic Key from Cancelable Fingerprint Templates", International Journal of Computer Science and Network Security, vol. 9, no.3, March 2009.

[23] Jang-Hee Yoo, Jong-Gook Ko, Sung-Uk Jung, Yun-Su Chung, Ki-Hyun Kim, Ki-Young Moon, and Kyoil Chung, "Design of an Embedded Multimodal Biometric System", ETRI-Information Security Research Division, 2007.

[24] Nageshkumar.M, Mahesh.PK and M.N. Shanmukha Swamy, "An Efficient Secure Multimodal Biometric Fusion Using Palmprint and Face Image", IJCSI International Journal of Computer Science Issues, Vol. 2, 2009.

[25] A.M. Bazen and S.H. Gerez, "Systematic methods for the computation of the directional fields and singular points of fingerprints", IEEE Transaction on Pattern Analysis and Machine Intelligence, vol. 24, no.7, pp.905–919, 2002.

[26] Yi Wang , Jiankun Hu and Fengling Han, "Enhanced gradient-based algorithm for the estimation of fingerprint orientation fields", Applied Mathematics and Computation, vol. 185, pp.823–833, 2007.

[27] Jinwei Gu and Jie Zhou, "A Novel Model for Orientation Field of Fingerprints", in Proceedings of the IEEE Computer Society Conference on Computer Vision and Pattern Recognition, vol.2, 2003.

[28] Keokanlaya Sihalath, Somsak Choomchuay, Shatoshi Wada and Kazuhiko Hamamoto, " Performance Evaluation Of Field Smoothing Filters", in Proceedings of  2th International  Conference on Biomedical Engineering (BMEiCON-2009), Phuket, Thailand, August 2009.

[29] D. Maltoni, D. Maio, A. K. Jain, and s. Prabhakar, Handbook of Fingerprint Recognition, Springer-Verlag, 2003.

[30] L. Hong, Y.Wan, and AI. Jain, "Fingerprint Image Enhancement: Algorithm and Performance Evaluation," IEEE Transactions on Pattern Analysis and Machine Intelligence, vol. 20, no. 8, pp. 777-789, August 1998.

[31] Feng Hao, Ross Anderson and John Daugman, "Combining Crypto with Biometrics Effectively", IEEE Transactions on Computers, vol. 55, no. 9, pp. 1081 - 1088, September 2006.

[32] Balasubramanian.K and Babu. P, "Extracting Minutiae from Fingerprint Images using Image Inversion and Bi-Histogram Equalization", Proceedings of SPIT-IEEE Colloquium and International Conference, Mumbai, India.

[33] M. Sepasian, W. Balachandran and C. Mares, "Image Enhancement for Fingerprint Minutiae-Based Algorithms Using CLAHE, Standard Deviation Analysis and Sliding Neighborhood", in Proceedings of the World Congress on Engineering and Computer Science 2008, San Francisco, USA, October 2008.

[34] Sharat Chikkerur, Alexander N. Cartwright and Venu Govindaraju, "Fingerprint enhancement using STFT analysis", Pattern Recognition, vol. 40, no.1, pp. 198-211, 2007.

[35] Greenberg, S.   Aladjem, M.   Kogan, D and Dimitrov, I, "Fingerprint image enhancement using filtering techniques" in Proceedings of the 15th International Conference on Pattern Recognition, vol.32, pp. 322-325, Barcelona, Spain, 2000.

[36] L. Lam, S. W. Lee, and C. Y. Suen, "Thinning Methodologies-A Comprehensive Survey", IEEE Transactions on Pattern analysis and machine intelligence, vol. 14, no. 9, 1992.

[37] Debnath Bhattacharyya, Poulami Das,Samir Kumar Bandyopadhyay and Tai-hoon Kim, "IRIS Texture Analysis and Feature Extraction for Biometric Pattern Recognition", International Journal of Database Theory and Application, vol. 1, no. 1, pp. 53-60, December 2008.

[38] J. Daugman, "Statistical Richness of Visual Phase Information: Update on Recognizing Persons by Iris Patterns," International Journal of Computer Vision, vol. 45, no. 1, pp. 25-38, 2001.

[39] S. Uma Maheswari, P. Anbalagan and T.Priya, " Efficient Iris Recognition through Improvement in Iris Segmentation Algorithm", International Journal on Graphics, Vision and Image Processing, vol. 8, no.2, pp. 29-35,  2008.

[40] John Daugman, "How Iris Recognition Works", in Proceedings of International Conference on Image Processing, vol.1, pp. I-33- I-36, 2002.

[41] David J. Field, "Relations between the statistics of natural images and the response properties of cortical cells", Journal of the Optical Society of America,vol.4, no. 12, 1987.

[42] Beng.A, Jin Teoh and Kar-Ann Toh, "Secure biometric-key generation with biometric helper", in proceedings of 3rd IEEE Conference on Industrial Electronics and Applications, pp.2145-2150, Singapore, June 2008.

[43] Sanaul Hoque , Michael Fairhurst and Gareth Howells, "Evaluating Biometric Encryption Key Generation Using Handwritten Signatures", in Proceedings of the 2008 Bio-inspired, Learning and Intelligent Systems for Security, pp.17-22, 2008.

[44] Gang Zheng, Wanqing Li and Ce Zhan, "Cryptographic Key Generation from Biometric Data Using Lattice Mapping", in Proceedings of the 18th International Conference on Pattern Recognition, vol.4, pp. 513 - 516, 2006.

[45] Tianhao Zhang, Xuelong Li, Dacheng Tao and Jie Yang, "Multimodal biometrics using geometry preserving projections", Pattern Recognition, vol. 41 ,  no. 3, pp. 805-813, March 2008.

[46] Donald E. Maurer and John P. Baker, "Fusing multimodal biometrics with quality estimates via a Bayesian belief network", Pattern Recognition, vol. 41, no. 3, pp. 821-832, March 2008.

[47] Muhammad Khurram Khana and Jiashu Zhanga, "Multimodal face and fingerprint biometrics authentication on space-limited tokens ", Neurocomputing, vol. 71, no. 13-15, pp.3026-3031, August 2008.

[48] Yan Yan and Yu-Jin Zhang , "Multimodal biometrics fusion using Correlation Filter Bank", in proceedings of the 9th International Conference on Pattern Recognition, pp. 1-4,Tampa, FL, 2008.

**Authors Detail:**

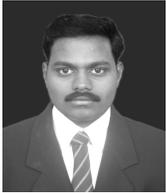

**Mr.A.Jagadeesan** was born in Coimbatore, India on June 14, 1979. He graduated from Bannari Amman Institute of Technology in 2000 with a degree in Electrical and Electronics Engineering. He completed his Master of Technology in Bio-medical Signal Processing and Instrumentation from SASTRA University in 2002. Thereafter he joined as a Lecturer in K.S.Rangasamy College of Technology till 2007. Now working as a Senior Lecturer in Bannari Amman Institute of Technology. He is a research scholar in the Department of Information and Communication Engineering. His area of interest includes Biometrics, Digital Image Processing, Embedded Systems and Computer Networks. He is a life member in ISTE and BMESI. He is also a member of Association of Computers, Electronics and Electrical Engineers (ACEE) and International Association of Engineers (IAENG).

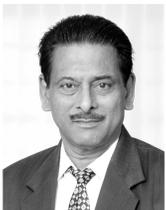

**Dr. K.Duraiswamy** received his B.E. degree in Electrical and Electronics Engineering from P.S.G. College of Technology, Coimbatore in 1965 and M.Sc. (Engg) from P.S.G. College of Technology, Coimbatore in 1968 and Ph.D. from Anna University in 1986. From 1965 to 1966 he was in Electricity Board. From 1968 to 1970 he was working in ACCET, Karaikudi. From 1970 to 1983, he was working in Government College of Engineering Salem. From 1983 to 1995, he was with Government College of Technology, Coimbatore as Professor. From 1995 to 2005 he was working as Principal at K.S.Rangasamy College of Technology, Tiruchengode and presently he is serving as Dean of KSRCT. He is interested in Digital Image Processing, Computer Architecture and Compiler Design. He received 7 years Long Service Gold Medal for NCC. He is a life member in ISTE, Senior member in IEEE and a member of CSI.